\documentstyle[prl,aps,graphics,epsfig]{revtex}
\begin{document}
\draft
\title{Quantum Computation with hot and cold ions:\\
An assessment of proposed schemes.}
\author{D. F. V. James\\
\small{Theoretical Division, Los Alamos National Laboratory, Los 
Alamos, NM 87545, USA} }
\date{\today}
\maketitle
%%%%%%%%%%%%%%%%%%%%%%%%%%%%%%%%%%%%%%%%%%%%%%%%%%%%%%%%%%%%%%%%%%%%%%%%%%%%%

% Abstract
\begin{abstract}
We present a brief critical review of the proposals for quantum 
computation with trapped ions, with particular emphasis on the
possibilities for quantum computation {\em without} the need for
cooling to the quantum ground state of the ions' collective oscillatory
modes.
\end{abstract}
%%%%%%%%%%%%%%%%%%%%%%%%%%%%%%%%%%%%%%%%%%%%%%%%%%%%%%%%%%%%%%%%%%%%%%%%%%%%%

%\bigskip
%PACS numbers:42.50.Vk, 32.80.Pj, 03.67.Lx, 42.25.Kb\\
\begin{center}
\bigskip
LA-UR-00-1436
\end{center}

%%%%%%%%%%%%%%%%%%%%%%%%%%%%%%%%%%%%%%%%%%%%%%%%%%%%%%%%%%%%%%%%%%%%%%%%%%%%%

%\pacs{PACS numbers:42.50.Vk, 32.80.Pj, 03.67.Lx, 42.25.Kb}
%%%%%%%%%%%%%%%%%%%%%%%%%%%%%%%%%%%%%%%%%%%%%%%%%%%%%%%%%%%%%%%%%%%%%%%%%%%%%

%\newpage

\section{Introduction}
Of all the proposed technologies for quantum information processing 
devices, argueably one of the most promising and certainly one of the
most popular is trapped ions.  This scheme, discovered by Ignacio Cirac 
and Peter Zoller \cite{cz}, and demonstrated experimentally shortly 
afterwards by Monroe et al. \cite{NISTgate}, is currently being 
persued by about a half-dozen groups world-wide \cite{groups}
(for an overview of this work, see, for example refs. \cite{SteaneRev, 
LANLrev,NISTrev}).

A vital ingredient of trapped ion quantum computing is the ability to
cool trapped ions down to their quantum ground state by sideband cooling. 
Using controlled laser pulses, the quantum state of the ion's collective
oscillation modes (i.e. the ions {\em external} degrees of freedom)
can then be altered conditionnally on the {\it 
internal} quantum state of the ions' valence electrons, and vice-versa. 
 This allows
quantum logic gates to performed.  The current state-of-the-art
(as of spring, 2000) is that two groups have suceeded in 
cooling strings of a few ions to the quantum ground state
\cite{NISTmodes,NISTdettang,BlattAdd,BlattEng},
and that entanglement of up to four ions has been experimentally
reported \cite{NISTfour}.

The fidelity of the quantum logic gates performed in trapped-ion
quantum computers
relies cruitially on the quantum state of the ions collective 
oscillatory degrees of freedom.  In the original Cirac-Zoller scheme
the ions must be in their quantum ground state of these degrees of 
freedom (the quanta of which are widely referred to as phonons).
If the purity of this quantum state
were to be degraded by the action of external perturbations (which, given
the fact that ions couple to any externally
applied electric field, seems quite likely) then the fidelity of 
quantum operations naturally will suffer.  The maintainance of the
cold ions in their oscillatory quantum ground state seems at
the moment to be the biggest single problem standing in the way of
advancing this field.  The solution is being tackled in two ways:
firstly the understanding and nuffication of the experimental causes
of the ``heating'' of the trapped ions, and secondly the investigation
of alternative schemes for performing quantum logic operations which
relax the strict condition of being in the quantum ground state of
the phonon modes.
This paper is a brief review and assessment of these schemes.

%%%%%%%%%%%%%%%%%%%%%%%%%%%%%%%%%%%%%%%%%%%%%%%%%%%%%%%%%%%%%%%%%%%%%%%%%
%%%%%%%%%%%%%%%%%%%%%%%%%%%%%%%%%%%%%%%%%%%%%%%%%%%%%%%%%%%%%%%%%%%%%%%%%
\section{Heating of ions}
The influence of random electromagnetic fields on trapped ions has
been analyzed by various authors 
\cite{Lamoreaux,Sara,Thomas,NISTheat,James:98a}; because this theory
impacts on our later discussions, we will give a brief reprise of it
here.
Consider $N$ ions confined in a trap.  The trap is
assumed to be sufficiently anistropic, and the ions sufficiently cold
that they lie crystalized along an axis of the trap in which the 
effective trapping potential is weakest, which we shall denote as the 
x-axis.  Because the ions are interacting via
the Coulomb force, their motion will be strongly coupled.  Their
small amplitude fluctuations are best described in terms of 
normal modes, each of which can be treated as an independent
harmonic oscillator \cite{James:98}.  There will be a total of $N$
such modes along the weak axis (we will neglect motion
along the directions of strong confinement).  We shall
number these modes in order of increasing resonance frequency,
the lowest ($p=1$) mode being the center of mass mode,
in which the ions oscillate as if rigidly clamped together. 
In the quantum mechanical description,
each mode is characterized by creation and annihilation operators
$\hat{a}^\dagger_{p}$ and $\hat{a}_{p}$ (where $p=1,\ldots N$).  The
ions are interacting with
an extrnal electric field ${\bf E}({\bf r},t)$. 
The Hamiltonian in this case is given by the expression
\begin{equation}
\hat{H}=i\hbar\sum^{N}_{p=1}\left[u_{p}(t)\hat{a}^\dagger_{p}-
u_{p}^{*}(t)\hat{a}_{p}\right],
\end{equation}
where
\begin{equation}
u_{p}(t)=\frac{i e}{\sqrt{2M\hbar\omega_{p}}}
\sum_{n =1}^{N}E_{x}({\bf r}_{n },t)b^{(p)}_{n }
\exp\left(i\omega_{p}t\right).\label{up}
\end{equation}
In eq.(\ref{up}), $b^{(p)}_{n}$ is the $n$-th element of the $p$-th
normalized eigenvector of the ion coupling matrix \cite{James:98},
$\omega_{p}$ being its resonance frequnecy, and $E_{x}$ is the
component of the electric field along the weak axis of the trap. 
In what follows, the center of mass phonon mode ($p=1$), whose
frequency is equal to the frequency $\omega_{x}$ of the Harmonic
trapping potential, will have special importance. 

The frequencies at which the externally applied fields are
resonant with the ions' motion is at most a few Megahertz;
the wavelengths of such radiation will therefore not be less
than 100 meters or so.  The separation of the ions is of the
order of 10 $\mu \mbox{m}$, or $10^{7}$ wavelengths. Thus 
spatial frequencies in the applied field on the spatial
scale of the ions' separation will be very evanescent, and to 
a very good approximation one can assume $E_{x}({\bf r}_{n },t) \approx
E_{x}(t)$, i.e. the field is constant over the extent of the ion
string.  Using the fact that $\sum_{n =1}^{N}b^{(p)}_{n 
}=\delta_{p,1}$, the interaction Hamiltonian becomes
\begin{equation}
\hat{H}=i\hbar u_{1}(t)\hat{a}^\dagger_{1} +h.a.,
\end{equation}
where
\begin{equation}
u_{1}(t)=\frac{i e}{\sqrt{2NM\hbar\omega_{x}}}E_{x}(t)
\exp\left(i\omega_{x}t\right),
\end{equation}
where $\omega_{x}\equiv\omega_{1}$ is the trapping frequency
along the x-axis.  In other words, spatially uniform fields
will only interact with the center-of-mass mode of the ions,
which is physical inituitive since some form of differential
force must be applied to excite modes in which ions move relative
to one another.

The dynamics governed by this Hamiltonian can be solved exactly
\cite{James:98a}.  The ``{\em heating time}', i.e. the time taken
for the occupation number of the center of mass mode to increase by
one, is given by the formula
\begin{equation}
\tau_{N}=\frac{M\hbar \omega_{x}}{N e^{2} E^{2}_{RMS} T},
\end{equation}
where $E_{RMS}$ is the root mean square value of $E_{x}(t)$ and
$T$ is its coherence time (we have assumed that $T \gg 
2\pi/\omega_{x}$).

\section{Quantum Computing using ``higher'' phonon modes}

The analysis of the heating of ions presented in the previous
section is directly linked to the first, and conceptually most simple
method for quantum computing with trapped ions in a manner which
avoids the heating problem  \cite{NISTmodes,James:98a}.  
Quite simply the ``higher'' ($p > 1$) modes of the ions' collective
oscillations can be utilized in place of the center of mass
($p=1$) mode originally considered by Cirac and Zoller.  The
pulse sequence required is exactly that proposed by those
authors, with the slight added complication that different
laser frequencies (i.e. the sideband corresponding the
stretch mode in question) must be employed, and that the laser-ion
coupling varies between different ions for the higher modes
\cite{James:98}, requiring different pulse durations for
different ions.
However, as has been pointed out by Saito
{\em et al.} \cite{Saito} (in the context of high-temperature
NMR experiments) the overall complexity of a computer 
algorithm involving {\em classical} control problems
of this kind can nullify any speed-up that
can be achieved via quantum parallelism.

Experimentally
the ``higher'' modes of the two-ion system are observed
to have heating times in excess of 5 $\mu$sec, as opposed
to heating times of less than 0.1 $\mu$sec for the
center of mass modes \cite{NISTmodes}, confirming that they are
indeed well isolated from the influence of external heating
fields, and can be used as a reliable quantum information bus.

The heating of the center of mass mode has an important
indirect influence.  As this mode becomes more and more
excited, the wavefunction of the ions becomes more 
spatially smeared-out, causing a random phase shift of the ions.
This effect is analogous to the Debye-Waller effect in X-Ray
crystallography \cite{NISTmodes}.
One possible solution for this problem has been proposed
\cite{Kielpinski}, namely the use of {\em sympathetic cooling}
by a separate species of ion, allowing the excitation
of the center of mass mode to be reduced and kept constant.
This scheme however poses the problem of devising a method
of loading a trap with an ion of a distinct species and
providing a second set of lasers to cool it.

%These ``higher'' modes are an example
%of a {\em decoherence free subspace} \cite{dfs}.  Such spaces are
%definied by their orthogonality to the decay operators appearing
%in the Lindblad form of the master equation.  

%%%%%%%%%%%%%%%%%%%%%%%%%%%%%%%%%%%%%%%%%%%%%%%%%%%%%%%%%%%%%%%%%%%%%%%%%
%%%%%%%%%%%%%%%%%%%%%%%%%%%%%%%%%%%%%%%%%%%%%%%%%%%%%%%%%%%%%%%%%%%%%%%%%

\section{Quantum computation with macroscopically resolved quantum 
states: the scheme of Poyatos, Cirac and Zoller}

The essential principle of the scheme proposed by Poyatos, Cirac and Zoller
\cite{Poyatos}
for ``hot'' ion quantum computation is to create coherent states of the
ions' collective oscillations, rather than Fock states.  A laser pulse,
appropriately tuned, flips the internal state of the ion and 
simultaneously provides a momentum ``kick'' to the wavepacket of 
the trapped ion {\em in a direction which is dependent on the internal
state of the ion}.  Thus if the ion/qubit is in state $|0\rangle$
it will start to move off in one direction; if it is in state $|1\rangle$
it will start to move in the opposite direction.  If it is in a 
superposition state, then a macroscopic entangled state (or ``cat'' 
state) will be created.  Because of the strong ion-ion coupling due to the Coulomb 
interaction, a second ion will also evolve into two spatially 
dependent wavepackets dependent on the state of the first ion (see 
Fig.\ref{fig1b}).  If the momentum kick imparted by the initial laser pulse
is sufficiently strong
then the wavepacket associated with the $|0\rangle$ will, after a 
short time, be spatially distinct from that associated with the $|1\rangle$
state. A laser may then be directed on that distinct wavepacket of the
second ion, allowing its state to be changed dependent on the state of
the first ion (Fig.\ref{fig1c}).  Once this is done, the motion of
the wavepackets in the traps restores them to their original positions
(Fig.\ref{fig1d}) and a third pulse, reversing the effect of the first pulse
and nullifying the momentum kick is applied to the first ion, 
completing the gate operation (Fig.\ref{fig1e}).

\begin{figure}[!ht]
\begin{center}
\epsfxsize=8.5cm  % \columnwidth
\epsfbox{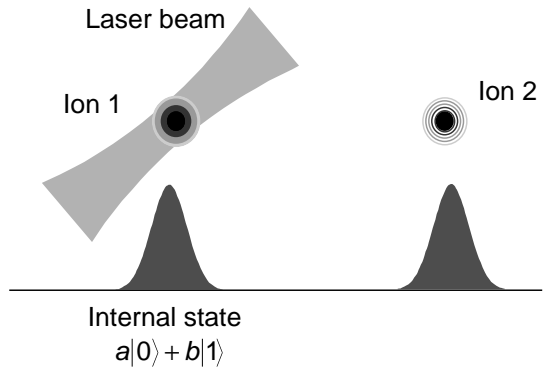}
\end{center}
\caption{Schematic picture of the spatial wavepackets
of the two trapped ion qubits interacting with the laser
to give a state-dependent momentum kick. }
\label{fig1a}
\end{figure}

\begin{figure}[!ht]
\begin{center}
\epsfxsize=8.5cm  % \columnwidth
\epsfbox{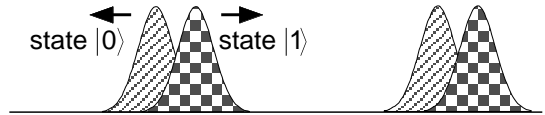}
\end{center}
\caption{As the wavefunction evolves in time, {\em both} ions'
wavepackets become spatially resolved dependent on the state
of the first ion.  The chequered wavepacket is associated with the
$|1\rangle$ state, the lined with the $|0\rangle$ state.}
\label{fig1b}
\end{figure}

\begin{figure}[!ht]
\begin{center}
\epsfxsize=8.5cm  % \columnwidth
\epsfbox{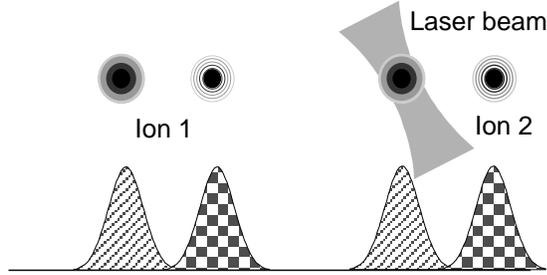}
\end{center}
\caption{When the two wavepackets are sufficiently separated,
a laser is used to flip the state of the second ion {\em dependent
on the state of the first ion}.  Note that, in practice, the 
separation of the wavepackets does not have to be greater than the
laser's spot size, provided that some controllable difference in
the illumination of the two packets is possible.}
\label{fig1c}
\end{figure}

\begin{figure}[!ht]
\begin{center}
\epsfxsize=8.5cm  % \columnwidth
\epsfbox{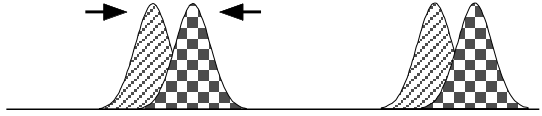}
\end{center}
\caption{After the flip, the wavepackets oscillate back to their 
initial spatial positions.  The wavepecket dynamics in this situation
is not simple harmonic motion, since the excitation pulse acts on all
of the phonon's modes.}
\label{fig1d}
\end{figure}

\begin{figure}[!ht]
\begin{center}
\epsfxsize=8.5cm  % \columnwidth
\epsfbox{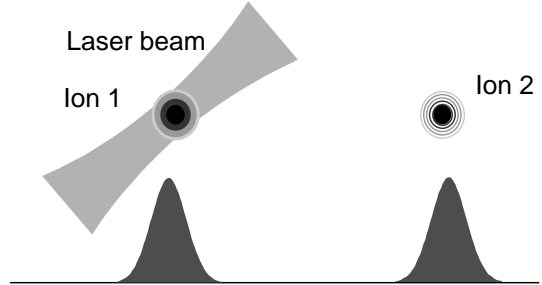}
\end{center}
\caption{Finally another laser pulse reverses the effect of the first 
pulse.}
\label{fig1e}
\end{figure}

The traveling wave laser pulses which provide the momentum kicks are described by
the interaction Hamiltonian
\begin{equation}
\hat{H}_{I}=\frac{\hbar\Omega}{2}\hat{\sigma}^{(+)}
\exp\left[i\sum_{p=1}^{N}\eta_{p}\left(\hat{a}_{p}+\hat{a}_{p}^{\dagger}\right)\right]+h.a.
\end{equation}
In this equation $\Omega$ is the Rabi frequency,  
which is proportional to the electric field strength of the laser
(see ref.\cite{James:98} for details) and the operators 
$\hat{\sigma}^{(+)} \equiv |0\rangle\langle 1|$ and
$\hat{\sigma}^{(-)} \equiv |1\rangle\langle 0|$ are respecitively the lowering
and raising operators for the internal states of the ion (treated
as a two level system).  In paper will be considering the dynamics of
one or two ions only, and the context should make it clear to which of 
the two ions the operators refer; in some cases subscripts are 
appended.  The constant $\eta_{p}$ is the Lamb-Dicke parameter, which
characterizes the strength of the coupling between the laser and the
oscillatory mode.  It varies between different modes and, in general,
from ion to ion.

If this Hamiltonian acts for a time $t_{las}=\pi/\Omega\ll 2\pi/\omega_{x}$
then the resultant transformation of the state of ion 1 is
\begin{equation}
|\phi'\rangle=
\left[\hat{\sigma}^{(+)}\prod^{N}_{p}\hat{D}_{p}\left(i\eta_{p}\right)+
\hat{\sigma}^{(-)}\prod^{N}_{p}\hat{D}_{p}\left(-i\eta_{p}\right)\right]|\phi\rangle ,
\end{equation}
where $\hat{D}_{p}\left(v\right)$ is the {\em displacement operator} for
the $p$-th phonon mode in question \cite{MandelWolf}.  The fact that all of
the phonon modes are excited by this operation leads to somewhat 
complicated dynamics of the excited wavepackets.  This is alieviated
somewhat by the use of slightly different trapping potentials 
(see \cite{Poyatos} for details), which may be realized in small
scale traps, in which the electrodes are close to the ions
(which could have a detrimental effect on the heating
of the ions.)
\footnote{Another possible method of modifying the ions' collective 
dynamics is to insert one or more ions of a different mass into the
ion chain, as has been investigated in the context of sympathetic
cooling by Kielpinski {\em et al.} \cite{Kielpinski}.}.  As the number
of ions increases
this dynamics will becomes more and more complicated, so that one
has to wait longer an longer times for the wavepackets to re-combine
(as in Fig.\ref{fig1d}) prior to completion of the gate.  This is
phenomnon unfortunately limits this scheme to no more than two or three
ions. 

In practice the ion wavepackets do not have to be {\em completely} 
separated spatially so that a laser can be focused on one but not the
other (as shown in Fig.\ref{fig1c}); so long as they
are separated somewhat, a laser
beam could be applied in such a fashion that one of the wavepackets
had its internal states flipped while the other recieved a pulse of
the same duration, but different intensity, contrived to leave the
internals states effectively unaltered (e.g. a ``$4\pi$'' pulse).
Care must however be exercised that laser fields are constant over
the spatial extent of each wavepacket, otherwise spatial information
will become inprinted on the internal degrees of freedom.

The finite temperature effects simply by increasing the size of the
ions' wavepackets. It derives its immunity from heating from
the use of macroscopic effects (i.e. the separation of the ions'
wavepackets) which are effected but slightly by the heating.
A possible source of decoherence would be differential heating, when
the two spatially separated wavepackets of a single ion
are excited by different random feilds, so that a mixed state 
of the internal degrees of freedom is 
created when the wavepackets are recombined.

%%%%%%%%%%%%%%%%%%%%%%%%%%%%%%%%%%%%%%%%%%%%%%%%%%%%%%%%%%%%%%%%%%%%%%%%%
%%%%%%%%%%%%%%%%%%%%%%%%%%%%%%%%%%%%%%%%%%%%%%%%%%%%%%%%%%%%%%%%%%%%%%%%%

\section{Quantum computation with virtual phonons: 
the scheme of M\o lmer and S\o rensen}

M\o lmer and S\o rensen have proposed related techniques
for creating both multi-ion entangled states \cite{molmer}
and for quantum computation \cite{sorensen1,sorensen2} with
ions in thermal motion.  The scheme proposed in fact is valid
for any mixed state of the ions' collective oscillation modes,
and is not confined to thermal equilibrium states. It relies on
the virtual excitation of phonon states, in a manner
analgous to the virtual excitation of some exited state of an atom
or molecule in Raman processes. Laser fields with two spectral
components detuned equally to the red and to the blue of the 
atomic resonance frequency are applied to a pair of ions in
the trap.  The interaction is described by the following
Hamiltonian

\begin{eqnarray}
\hat{H}_{I}&=&\hbar\Omega\hat{J}^{(+)}
\left\{1+i\eta
\left(
\hat{a}e^{-i\omega_{x}t}+\hat{a}^{\dagger}e^{i\omega_{x}t}\right)
\right\}\cos(\delta t)+
h.a. \nonumber\\
&=&\hbar\Omega e^{i\delta t}\hat{J}_{x}-
\hbar\Omega\eta
e^{i(\delta+\omega_{x})t}\hat{a}^{\dagger}\hat{J}_{y}-
\hbar\Omega\eta
e^{i(\delta-\omega_{x})t}\hat{J}_{y}\hat{a}+ h.a.
\end{eqnarray}
In this equation $\delta$ is the detuning of the laser beam from 
the resonance frequnecy of the two level system.  For large values
of $\delta$ it is convenient to consider this interaction in terms of
an effective Hamiltonian (see appendix), which neglects the effects
of very rapidly varying terms.  In this case, the effective Hamiltonian
is
\begin{eqnarray}
\hat{H}_{eff}&=& 
\frac{\hbar\Omega^{2}\eta^{2}}{(\delta+\omega_{x})}
\left[\hat{J}_{y}\hat{a}, \hat{a}^{\dagger}\hat{J}_{y}\right]+
\frac{\hbar\Omega^{2}\eta^{2}}{(\delta-\omega_{x})}
\left[\hat{a}^{\dagger}\hat{J}_{y},\hat{J}_{y}\hat{a}\right] 
\nonumber\\
&=&\frac{\hbar\Omega^{2}\eta^{2}}{(\delta-\omega_{x})}
\left(\frac{2\omega_{x}}{\delta+\omega_{x}}\right)
\hat{J}^{2}_{y}.
\end{eqnarray}
This interaction is equivalent to a conditional quantum logic gate
preformed between the two ions, and can be used to create 
multiparticle entangled states. 

This scheme is very attractive because, while it has the
possibility of being scalable to many ions, its operation is
{\em independent} of the occupation number of the phonon modes,
and so its fidelity is not degraded by excitation {\em during}
the gate operations themselves.
Its chief drawback seems to be the time taken to perform
gate operations. In \cite{sorensen1} an example is given of population
oscillations associated with the above entangling operations in the
presence of noise.  The Rabi frequency of these oscillations was
approximately 4500$\omega_{x}$ (c.f. Fig.4 of ref.\cite{sorensen1},
with appropriate change of notation).  Given that trap frequencies
must be of the order of $\omega_{x}\sim (2\pi)$500 kHz in order for
the ions to be individually resolvable by focused lasers
\footnote{It is not necessary to
resolve ions individually for this scheme to be used to create
entanglement; however some form of differential laser addressing
will be necessary in order to perform
quantum computations involving more than two qubits.}
, this implies
a gate time of the order of 50 milliseconds.  As explained in
ref.\cite{sorensen2}, it is possible to decrease this time by reducing the
detuning $\delta$ of the laser, at the cost of increasing the 
susceptibility of this scheme to heating during gate operations.
Nevertheless M\o lmer and S\o rensen's scheme is a very compelling
idea, and has been used experimentally to create entangled
states of multiple ions \cite{NISTfour}.

%%%%%%%%%%%%%%%%%%%%%%%%%%%%%%%%%%%%%%%%%%%%%%%%%%%%%%%%%%%%%%%%%%%%%%%%%
%%%%%%%%%%%%%%%%%%%%%%%%%%%%%%%%%%%%%%%%%%%%%%%%%%%%%%%%%%%%%%%%%%%%%%%%%

\section{Quantum computation via adiabatic passages: the scheme of 
Schneider, James and Milburn}

The scheme proposed by  Schneider et al. \cite{Schneider} relies on
two operations: first the phonon-number dependent a.c. Stark shift
introduced by D'Helon and Milburn \cite{D'Helon96}, and second the
use of stimulated Raman adiabatic passage to carry out certain kinds
of transitions independently of the occupation number of the phonon
mode used as a quantum information bus.

First let us consider the origin of the D'Helon-Milburn shift.  
The Hamiltonian for a single two-level ion at the node of a detuned classical 
standing wave is given by the following formula:

\begin{eqnarray}
\hat{H}_{I}&=&\frac{\hbar\Omega\eta}{2}\hat{\sigma}^{(+)}
\left(\hat{a}e^{-i\omega_{x}t}+\hat{a}^{\dagger}e^{i\omega_{x}t}\right)
e^{i\Delta t}+h.a. \nonumber\\
&=&\frac{\hbar\Omega\eta}{2}
\left(\hat{\sigma}^{(+)}\hat{a}e^{i(\Delta-\omega_{x})t}+
\hat{\sigma}^{(+)}\hat{a}^{\dagger}e^{i(\Delta+\omega_{x})t}\right)
+h.a., \nonumber\\
\end{eqnarray}
where $\Delta$ is the laser detuning.
In the limit of large detuning ($\Delta \gg \omega_{x}$)  the
{\em effective} Hamiltonian is (using the result derived in
the appendix):

\begin{eqnarray}
\hat{H}_{eff}&=& 
\frac{\hbar\Omega^{2}\eta^{2}}{2(\Delta-\omega_{x})}
\left[\hat{\sigma}^{(-)}\hat{a}^{\dagger}, \hat{\sigma}^{(+)}\hat{a}\right]+
\frac{\hbar\Omega^{2}\eta^{2}}{2(\Delta+\omega_{x})}
\left[\hat{\sigma}^{(-)}\hat{a}, \hat{\sigma}^{(+)}\hat{a}^{\dagger}\right] 
\nonumber\\
&\approx&-\frac{\hbar\Omega^{2}\eta^{2}}{2\Delta}
\left(2\hat{n}+1\right) \hat{\sigma}_{z}=
-\frac{\hbar\Omega^{2}\eta^{2}}{\Delta}\hat{n}\left(\hat{\sigma}^{(+)}+1/2\right)
+\frac{\hbar\Omega^{2}\eta^{2}}{2\Delta}\left(\hat{n}-\hat{\sigma}^{(+)}\right).
\label{heffsara}
\end{eqnarray}
The second term on the right hand side of the final equation 
represents a level shift, which can be compensated for by detuning 
the laser.
If we choose the duration $\tau$ of this interaction to be 
$\tau = \pi \Delta/\Omega^{2}\eta^{2} $, the time evolution is represented by the 
operator  
\begin{equation}
\hat{\cal{S}}_{t} = \exp[-i\hat{a}^\dagger\hat{a} (\hat{\sigma}_z + 1/2)\pi] \,\, .
\end{equation} 
This time evolution flips the phase of the ion
when the CM mode is in an odd state and the ion is in its 
excited state, thus providing us with a conditional phase shift 
for an ion and the CM mode.   This operation will be performed only
on one of the ions (the target qubit) involved in the quantum gate
(which we denote by the subscript $t$).  Operations acting on the 
second ion involved in the gate (the control qubit) will be denoted
by the subscript $c$.

\begin{center}
\begin{figure}[!ht]
\vskip 2cm
\epsfxsize=8.5cm
\epsffile{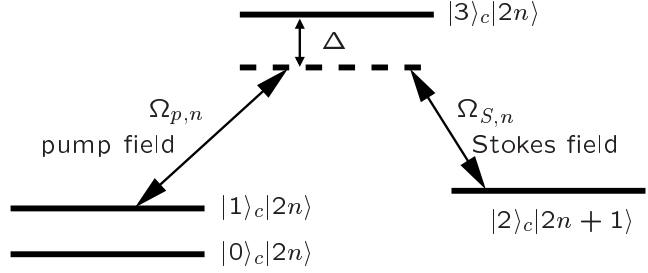}
\vskip 1cm
\caption{Illustration of the level scheme of the
control ion used to realize the adiabatic passage operations
${\cal{A}}^{+}_{c}$ and ${\cal{A}}^{-}_{c}$.}
\label{adpass}
\end{figure}
\end{center}

The adiabatic passage \cite{Bergmann95} required for the gate operation can be 
realized using two lasers, traditionally called the pump and 
the Stokes (see fig.\ref{adpass}). The 
pump laser is polarized to couple the control qubit state $|1\rangle_c$ to 
some second auxiliary state
$|3\rangle_c$ and is detuned by an amount $\Delta$.
The Stokes laser couples to the red side band transition $|2\rangle_c
|n+1\rangle \leftrightarrow |3\rangle_c |n\rangle$, with the
same detuning $\Delta$. If the population we want to transfer 
adiabatically is 
initially in the state $|1\rangle_c |n\rangle$, 
we turn on the
Stokes field (i.e.~the sideband laser) and then slowly turn on 
the pump field (i.e.~the carrier laser) until both lasers are turned 
on fully. Then we slowly 
turn off the Stokes laser: this is the famous ``counter-intuitive'' 
pulse sequence used in adiabatic passage techniques.
The adiabatic passage must be performed very 
slowly. The condition in our scheme is that 
$T \gg 1/\Omega_{p,n}, 1/\Omega_{S,n}$, where $T$ is the duration of 
the adiabatic passage and $\Omega_{p,n}$ ($\Omega_{S,n}$) are the 
effective Rabi frequencies for the pump and the Stokes transition, 
respectively \cite{QM}.  
Using the adiabatic passage we can transfer the population from 
$|1\rangle_c |n\rangle$ to $|2\rangle_c |n+1\rangle$. To 
invert the adiabatic passage, we just have to interchange the roles 
of the pump and the Stokes field. We will denote the
adiabatic passage by operators ${\cal{A}}^{+}_{1}$ and 
${\cal{A}}^{-}_{1}$
defined as follows:
\begin{eqnarray}
{\cal{A}}_j^+&:&|1\rangle_j |n\rangle \rightarrow  |2\rangle_j 
|n+1\rangle\nonumber \\
{\cal{A}}_j^-&:&|2\rangle_j |n+1\rangle \rightarrow  
|1\rangle_j|n\rangle\,\, .
\end{eqnarray}

The utility of this adiabatic passage scheme is that,
despite the fact that the laser trasnition rates $\Omega_{p,n}$ 
and $\Omega_{S,n}$ are dependent on the phonon occupation number $n$,
the adiabatic passage using the counter-intuitive pulse sequence
is {\em independent} of $n$. 

\begin{center}
\begin{figure}[!ht]
\vskip 1cm
\epsfxsize=8.5cm
\epsffile{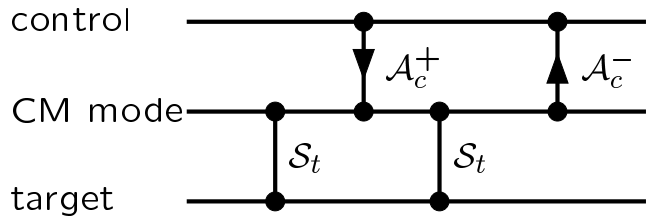}
\vskip 1cm
\caption{Schematic illustration of the steps
involved in the CROT gate with hot ions. The individual steps are
discussed in detail in the text.}
\label{saragate}
\end{figure}
\end{center}

These two operations are combined in the sequence shown in 
fig.\ref{saragate} in order to perform quantum gate operations.
A detailed breakdown of the operation, including the intermediate
states are very stage, is given in \cite{Schneider}. 

This principle drawbacks of this scheme are two-fold
\footnote {In order that the following remarks
be view in their correct context, the reader should be aware that
the author of the present article was one
of the authors of the scheme by Schneider {\em et al.}}.  Because of
the adiabatic passage involved, it will of necessity be slow, gates
requiring times of the order of a milliseconds
(although it has this in common with the other schemes described
in this paper.)  Secondly this scheme (unlike those of Poyatos {\em et 
al.} and M\o lmer and S\o rensen) is vulnernable to heating {\em 
during} the gate operation.  

A further complication is the presence
of multiple phonon modes in real experiments; to take into account
their influence, eq.(\ref{heffsara}) needs to be rewritten as
follows:

\begin{equation}
\hat{H}_{eff}=-\frac{\hbar\Omega^{2}\eta^{2}}{2\Delta}\sum_{p=1}^{N}
\eta^{2}_{p}\left(2\hat{n}_{p}+1\right) \hat{\sigma}_{z}.
\label{heffsara2}
\end{equation}

The $\hat{\cal{S}}_{t}$ gate will only function as designed when all of
the modes except the one to be used as the quantum information bus have
zero population.  Thus this scheme at best is a means of avoiding the
necessity of reducing the population of {\em every} mode to its
quantum ground state; one mode can be left in a mixed state.

%%%%%%%%%%%%%%%%%%%%%%%%%%%%%%%%%%%%%%%%%%%%%%%%%%%%%%%%%%%%%%%%%%%%%%%%%
%%%%%%%%%%%%%%%%%%%%%%%%%%%%%%%%%%%%%%%%%%%%%%%%%%%%%%%%%%%%%%%%%%%%%%%%%
%
%\section{Quantum computation via non-communting operators: the scheme of 
%Milburn}

%\begin{equation}
%\hat{H}_{I}=i\hbar\left(\alpha\hat{a}^{\dagger}-\alpha^{\ast}\hat{a}\right)\hat{J}_{z}

%\end{equation}

%$\hat{J}_{z}=\left(\hat{\sigma}_{c,z}+\hat{\sigma}_{t,z}\right)$

%%%%%%%%%%%%%%%%%%%%%%%%%%%%%%%%%%%%%%%%%%%%%%%%%%%%%%%%%%%%%%%%%%%%%%%%%
%%%%%%%%%%%%%%%%%%%%%%%%%%%%%%%%%%%%%%%%%%%%%%%%%%%%%%%%%%%%%%%%%%%%%%%%%
\section{Assessment}

The various schemes for quantum computation with trapped ions
in principle meet many of
the criteria for scalable quantum computation technology.  Here we 
discuss the various criteria one by one.

\subsection{Initialization}
The quantum information register (the ions) and
the quantum information bus (the phonon modes) can be initialized
reliably using laser cooling and optical pumping.  
Important aspects of these techniques have already been
demonstrated experimentally.  The ``hot'' ion schemes discussed
here, if they can be realized experimentally, ease the stringent
requirements on preparation of the initial state of the ions
collective oscillation modes.

\subsection{Gate Operations}
Quantum logic can be performed using the various schemes outlined
above.  The common ingredient is laser control of the quantum states
of the ions internal and external degrees of freedom, requiring
pulses of known duration and strength focused acurately on individual
ions.  Methods for aleiviating the laser focusing problem by altering the
ions resonance frequency by various means such as non-uniform electric
or magnetic fields have been proposed \cite{NISTrev,Leibfried}.  
The ability to address individual ions with laser beams and control their quantum
states has been demonstrated experimentally by two groups using various
means \cite{NISTdettang,BlattAdd}.

\subsection{Isolation from the Environment}
The internal degrees of freedom of the ions, in
which the quantum information is stored, have very long
decoherence times (especially when Raman transitions
form the basis of the single-qubit operations.)   The
principal form of enviromental disrupion suffered by ion
traps is disturbance of the motional degrees of freedom,
the proposed methods of avoiding this problem being the
subject of this article.

The ``higher modes'' scheme is well isolated from the environment,
except for the indirect influence of the Debye-Waller effect.
Both the Poyatos {\em et al.} scheme and the M\o lmer-S\o rensen scheme
are not intrinsically isolated from the environment, but
avoid its influence in various ingenious ways. The Schneider {\em et al.}
will suffer from environmental influences during gate operations
unless they can be nullified, for example by using ``higher modes''.

\subsection{Error correction}
There is nothing instrinsic that will rule out implimentation
of fault tolerant quantum computation in ion traps when sufficient
numbers of ions become availible.  Ancilla ions can be prepared in
their quantum ground state independent of other ions in the register.
The use of multiple stretch modes (there are N-1 such modes in the
weak trapping direction), allows quantum gates to be performed in
parallel.  Read out can be performed at intermediate stages during
calculations without destroying the qubit being read, or disturbing
other ions in the register unduly (there will be recoil during the
read out that has the possibility of excitation of the oscillatory 
modes).

\subsection{Read Out}
The read-out of the quantum state of ions using a cycling transition
has been demonstrated experimentally with high efficiency and 
reliability \cite{NISTgate}. Indeed these experiments are the
{\em only} ones in which high efficiency strong measurement of
a single quantum system (as opposed to an ensemble of systems)
has been performed.

\subsection{Scalability}
The  ultimate number of ions that can be stored in a string in an ion trap and 
used for quantum computation
is limited by a number of factors.  Probably the most important is
growing complexity of the sideband spectrum as the number of ions
grows.  Even in the case of highly anisotropic traps (in which
transverse oscillations can be neglected) the number of oscillation
modes is equal to the number of ions, and each mode has a distinct
frequency, with an infinite ladder of excitation resonances.  In
addition one has to take into account multi-phonon resonances;
the whole leading very complicated structure in frequency space.
The extent to which this ``spectrum of death'' \cite{SoD} can be understood
and exploited, by systematic identication of
resonances, careful bookeeping and tayloring of Lamb-Dicke coefficients
remains to be seen.  Other effects which place an upper bound on the
number of ions in a single register is that fact that the spatial
separation of the ions decreases $\propto N^{-0.56}$ \cite{meyrath}, making their
spatial resolution by a focused laser beam more and more difficult.
Another, more definite upper bound on the 
number of ions that can be stored in a linear configuration
is the onset of a phase transition to a more complex configuration
such as a zig-zag \cite{Enzer:00}; for traps optimized for quantum
computation with singly ionized calcium
this occurs at about 170 ions.  It is however argueable whether or
not the onset of instabilities makes quantum computation impossible.

If only small numbers of ions can be reliably used for quantum 
computation in a single ion trap, multiple traps will be needed for
large scale devices.  DeVoe \cite{DeVoeArrays} has proposed fabricating
multiple elliptical traps, each suitable for a few dozen ions, on a
substrate with a density of 100 traps/cm$^{2}$.  Some form of reliable,
high efficiency quantum communication channel to link the multiple 
traps would need to be implemented 
\cite{QuantLink,QuantLinkSin,QuantLinkUs}.
An alternative scheme has been proposed by Wineland {\em et al.}
\cite{NISTrev,Nisttwotraps} in which two traps are used.  One
trap is used to store a large number of ions in a readily
accessible manner (e.g. in an easily rotated ring configuration); 
each of these ions form the
qubits of the register of the quantum computer.  When a gate
operation is to be performed, the two involved ions are
extracted from the storage trap by applying static electric
fields in an appropriate controlled manner, and transfered to
a separate logic trap where they can be cooled and quantum logic operations
can be performed on them.  The cooling could be done sympathetically by
a third ion of a separate species stored in the logic trap (thereby
preserving the quantum information stored in the two
logic ions which otherwise would be lost during cooling); in these 
circumstances either the original Cirac-Zoller
scheme or any of the ``hot gates'' schemes described here can be
used as the mechanism for peforming the logic; in particular the
Poyatos {\em et al.} scheme, whose principal drawback seems to be
its lack of scalability beyond two or three ions, would no longer
be at a disadvantage, and given that it is considerably faster than
both the M\o lmer-S\o rensen and Schneider {\em et al.} schemes, might
be attactive.

In conclusion, the variety and richness of the quantum computing schemes that
have been devised for ion traps illustrates the great flexibility of
this technology.  Uniquely amoungst the proposals for quantum 
computing technology, the question for
ion traps is not ``does it work?'' but rather ``how far can it be developed?''

\section*{Acknolwledgements}
The author wishes to thank Gerard Milburn, Sara Schneider, Andrew 
White, Michael Holzscheiter and Dave Wineland for useful conversations
and correspondence.  This work was performed in part while the author
was a guest at the Deptartment of Physics, University of Queensland,
Brisbane, Australia.  He would like to thank the faculty, staff and
students for their warm hospitality.  This work was funded in part
by the U.S. National Security Agency.
%%%%%%%%%%%%%%%%%%%%%%%%%%%%%%%%%%%%%%%%%%%%%%%%%%%%%%%%%%%%%%%%%%%%%%%%%
%%%%%%%%%%%%%%%%%%%%%%%%%%%%%%%%%%%%%%%%%%%%%%%%%%%%%%%%%%%%%%%%%%%%%%%%%

\section*{Appendix: Effective Hamiltonians for Detuned Interactions}

We start with the Schr\"odinger equation in the interaction picture, 
i.e.
\begin{equation}
i\hbar \frac{\partial}{\partial t}|\psi(t)\rangle=\hat{H}_{I}(t)|\psi(t)\rangle
\label{SchroEqnApp}
\end{equation}
The {\em formal} solution of this first order partial differential 
equation is
\begin{equation}
|\psi(t)\rangle=|\psi(0)\rangle+\frac{1}{i \hbar}\int^{t}_{0} 
\hat{H}_{I}(t')|\psi(t')\rangle dt'.
\end{equation}
Subtituting this result back into eq.(\ref{SchroEqnApp}), we 
obtain
\begin{equation}
i\hbar \frac{\partial}{\partial t}|\psi(t)\rangle=\hat{H}_{I}(t)|\psi(0)\rangle+
\frac{1}{i \hbar}\int^{t}_{0} 
\hat{H}_{I}(t) \hat{H}_{I}(t')|\psi(t')\rangle dt'
\label{SchroEqnAppB}
\end{equation}
If we assume that the interaction Hamiltonian is strongly detuned, in 
the sense that $\hat{H}_{I}(t)$ consists of a number of highly 
oscillative terms, then to a good approximation the first term
on the right hand side of eq.(\ref{SchroEqnAppB}) can be neglected,
and we can adopt a Markovian approximation for the second term, 
so that the evolution of $|\psi(t)\rangle$ is approximately
governed by the following equation
\begin{equation}
i\hbar \frac{\partial}{\partial t}|\psi(t)\rangle \approx
\hat{H}_{eff}(t)|\psi(t)\rangle ,
\label{SchroEqnAppC}
\end{equation}
where
\begin{equation}
\hat{H}_{eff}(t) = \frac{1}{i \hbar}\hat{H}_{I}(t)
\int \hat{H}_{I}(t') dt',
\label{EffHamApp}
\end{equation}
where the indefinite integral is evaulated at time $t$ without a constant
of integration. These arguments can be placed on more rigorous footing by
considering the evolution of a time-averaged wavefunction.

We will now assume that the interaction Hamilitonian consists explicitly
of a combination of harmonic time varying components, i.e.
\begin{equation}
\hat{H}_{I}(t)=\sum_{m}\hat{h}_{m} \exp(i\omega_{m}t)+h.a.,
\end{equation}
where $h.a.$ stands for the the hermitician adjoint of the
perceeding term, and the frequencies $\omega_{m}$ are all
distinct (i.e. $m\neq n \Leftrightarrow \omega_{m} \neq \omega_{n}$).
In this case the effective Hamiltonian $\hat{H}_{eff}(t)$ reduces
to a simple form useful in the analysis of laser-ion interactions:
\begin{eqnarray}
\hat{H}_{eff}(t) &=& \sum_{m,n} \frac{1}{i\hbar}
\left(\hat{h}_{m} e^{i\omega_{m}t}+\hat{h}^{\dagger}_{m} e^{-i\omega_{m}t}\right)
\left(\hat{h}_{n} \frac{e^{i\omega_{n}t}}{i\omega_{n}}+\hat{h}^{\dagger}_{n} 
\frac{e^{-i\omega_{n}t}}{-i\omega_{n}}\right) \nonumber \\
&=&\sum_{m,n} \frac{1}{-\hbar \omega_{n}} 
\left(\hat{h}_{m}\hat{h}_{n} e^{i(\omega_{m}+\omega_{n})t}+
\hat{h}_{m}\hat{h}^{\dagger}_{n} e^{i(\omega_{m}-\omega_{n})t}-
\hat{h}_{m}^{\dagger}\hat{h}_{n} e^{-i(\omega_{m}-\omega_{n})t}-
\hat{h}_{m}^{\dagger}\hat{h}^{\dagger}_{n} e^{-i(\omega_{m}+\omega_{n})t}
\right) \nonumber \\
&=&\sum_{m}\frac{1}{\hbar\omega_{m}} 
[\hat{h}^{\dagger}_{m},\hat{h}_{m}]+ 
\mbox{oscillating terms} .
\end{eqnarray}

If we confine our interest to dynamics which are time-averaged over a 
period much longer than the period of any of the oscillations present
in the effect Hamiltonian (i.e.  averaged over a time $T \gg 
2\pi/\min\{|\omega_{m}-\omega_{n}|\}$) then the oscillating terms may 
be neglected, and we are left with the following simple formula for 
the effective Hamiltonian:

\begin{equation}
\hat{H}_{eff}(t) = \sum_{m}\frac{1}{\hbar\omega_{m}} 
[\hat{h}^{\dagger}_{m},\hat{h}_{m}] .
\end{equation}

%%%%%%%%%%%%%%%%%%%%%%%%%%%%%%%%%%%%%%%%%%%%%%%%%%%%%%%%%%%%%%%%%%%%%%%%%%%%%


\begin{thebibliography}{10}
\bibitem{SoD} This evocative phrase was introducted by C.~Monroe in public 
lectures to describe the complex side-band spectrum of trapped
Be ions.
\end{thebibliography}
\end{document}